\RequirePackage[2018-12-01]{latexrelease}
\documentclass[showpacs,showkeys,11pt,
preprint,preprintnumbers,nofootinbib,
groupedaddress,superscriptaddress,amsmath,amssymb]{revtex4}
\usepackage{cancel}
\usepackage{xcolor}
\usepackage{tabularx}
\usepackage{amsfonts}
\usepackage{amsmath}
\usepackage{graphicx,subfigure}
\usepackage{caption}
\usepackage[export]{adjustbox}
\usepackage{verbatim} 
\usepackage{epsfig}
\usepackage{epstopdf}
\usepackage{url}
\usepackage{multirow}
\usepackage{hhline}
\usepackage{feynmp}
\usepackage{booktabs}
\usepackage{csquotes}

\newcommand {\be}{\begin{equation}}
\newcommand {\ee}{\end{equation}}
\newcommand {\ba}{\begin{eqnarray}}
\newcommand {\ea}{\end{eqnarray}}

\newcommand {\cbma}{c_{\beta-\alpha}}

\newcommand {\tb}{t_{\beta}}

\begin{document}

\title{Machine Learning in the Hunt for Charged Higgs Bosons at Gamma-Gamma Colliders in the Type III Two Higgs Doublet Model}
\pacs{12.60.Fr, 
      14.80.Fd  
}
\keywords{Charged Higgs, 2HDM-III, Photon--Photon Collider, ILC, CLIC, Multivariate Analysis, Neural Networks}

\author{Abdul Quddus}
\email{abdulqudduskakakhail@gmail.com}
\affiliation{Federal Urdu University of Arts, Science and Technology, Islamabad, Pakistan}

\author{Ijaz Ahmed}
\email{ijaz.ahmed@fuuast.edu.pk}

\author{Jamil Muhammad}
\email{mjamil@konkuk.ac.kr}
\affiliation{Sang-Ho College, and Department of Physics, Konkuk University, Seoul 05029, South Korea}
\author{M. A. Arroyo-Ure\~na}
\email{marco.arroyo@fcfm.buap.mx}
\affiliation{Facultad de Ciencias F\'isico-Matem\'aticas and
Centro Interdisciplinario de Investigaci\'on y Ense\~nanza de la Ciencia (CIIEC), Benem\'erita Universidad Aut\'onoma de Puebla, C.P. 72570, Puebla, Pue., M\'exico}

\author{Usman Ahmad}
\affiliation{Federal Urdu University of Arts, Science and Technology, Islamabad, Pakistan}

\begin{abstract}
We study charged Higgs boson pair production, $\gamma\gamma\to H^{+}H^{-}$, in the Type-III Two-Higgs-Doublet Model (2HDM-III) for $100\leq M_{H^{\pm}}\leq190~\text{GeV}$ at a $\gamma\gamma$ collider with $\sqrt{s}=3~\text{TeV}$ and $\mathcal{L}_{int}=3000~\text{fb}^{-1}$. Two benchmark points, BP1 ($\tan\beta=10$, $\cos(\beta-\alpha)=-0.2$) and BP2 ($\tan\beta=20$, $\cos(\beta-\alpha)=-0.1$), are chosen in agreement with theoretical constraints, flavor-physics data and the signal strengths of the $125~\text{GeV}$ Higgs boson. The total width of $H^{\pm}$ is $10^{-4}$--$10^{-1}~\text{GeV}$, i.e. a narrow resonance, and $\mathrm{BR}(H^{+}\to\tau\nu_{\tau})$ falls from $4.6\times10^{-3}$ (BP1) and $2.9\times10^{-4}$ (BP2) at $100~\text{GeV}$ to $1.8\times10^{-4}$ and $1.2\times10^{-5}$ at $190~\text{GeV}$. The $\tau\nu_{\tau}$ and $tb$ final states are analyzed with four TMVA classifiers (BDT, MLP, Likelihood and LikelihoodD) using $1.0\times10^{5}$ generated events per process. For $171$ signal and $604$ background events, the MLP and BDT reach AUC values of $0.934$ and $0.933$ and maximum significances $S/\sqrt{S+B}$ of $9.86$ and $9.82$, compared with $0.870$ and $8.54$ for LikelihoodD and $0.815$ and $7.55$ for Likelihood. The preselection cuts raise the significance by only $0.09$--$0.91\%$, depending on the classifier. The quoted significances are statistical; systematic uncertainties are not included.
\end{abstract}
\maketitle
\section{Introduction }
The 2012 observation of a scalar resonance near $125~\text{GeV}$ by ATLAS and CMS at the Large Hadron Collider (LHC) \cite{atlas2012observation, ATLAS:2022map, CMS:2022portrait, PDG:2024} completed the particle content of the Standard Model (SM), and its measured couplings agree with the SM expectation so far. The SM scalar sector consists of a single doublet, which, through the Brout--Englert--Higgs mechanism, breaks the electroweak symmetry (EWSB) and generates the gauge-boson masses; it therefore contains no electrically charged scalar \cite{Dutta:2025}. Many beyond-the-SM constructions enlarge the scalar sector and thereby predict charged Higgs bosons: the Two-Higgs-Doublet Model (2HDM), supersymmetry, composite Higgs scenarios, the Littlest Higgs model \cite{Logan:2004}, doubly charged Higgs scenarios \cite{Chakrabarti:1998}, grand unified theories and axion models. Among these, the 2HDM is a common building block or low-energy limit of several others, for example supersymmetric scenarios such as the Minimal Supersymmetric Standard Model (MSSM) and composite-Higgs constructions, which makes it a convenient framework for phenomenological studies. The production and decay properties of $H^{\pm}$ depend on how the two doublets couple to quarks and leptons, and observing such a state would be unambiguous evidence for an extended Higgs sector.

Linear colliders can be operated in $e^{+}e^{-}$ and $\gamma\gamma$ modes, and the photon mode of the ILC and CLIC is attractive for charged-scalar searches \cite{Bambade:2019, Narain:2022, Brunner:2022, Adli:2013}: energetic photons derived from the lepton beams collide directly, which opens access to a variety of processes, including $H^{+}H^{-}$ production \cite{Asner:2003, Borden:1993, Gunion:1993}. In $e^{+}e^{-}\rightarrow H^{+}H^{-}$ the pair is produced through $s$-channel $\gamma^{*}/Z^{*}$ exchange, whose contribution decreases as the energy grows, whereas $\gamma\gamma\rightarrow H^{+}H^{-}$ is not affected by this suppression; earlier studies therefore indicate that the $\gamma\gamma$ rate may exceed the $e^{+}e^{-}$ one at high energies \cite{Ouazghour:2024, Asakawa:2010}. Although both modes have been computed at various orders, how higher-order corrections modify the predictions in the various models remains to be explored in full.

The present paper builds on our earlier multivariate study of $\gamma\gamma\rightarrow H^{+}H^{-}$ at $\sqrt{s}=3~\text{TeV}$ \cite{Ahmed:2025}, which addressed heavy charged Higgs bosons in Type-I and Type-II 2HDM benchmark scenarios. Here we turn to the Type-III model and to the intermediate mass window $100~\text{GeV}\leq M_{H^{\pm}}\leq 190~\text{GeV}$, a region that remains difficult at the LHC because of large SM backgrounds \cite{Laurila:2019, ATLAS:2021hplus}. The benchmark points, the event selection and the set of classifiers differ from those of Ref.~\cite{Ahmed:2025}; in particular, the plain Likelihood classifier is added to the BDT, MLP and LikelihoodD. For benchmark points (BPs) that satisfy the theoretical and flavor-physics constraints, we study the $H^{\pm}\to\tau\nu_{\tau}$ and $H^{\pm}\to tb$ channels at a $\gamma\gamma$ collider with $\sqrt{s}=3.0~\text{TeV}$ and $\mathcal{L}_{int}=3000~\text{fb}^{-1}$. The BDT and MLP are compared with likelihood-based classifiers, with the aim of showing how machine-learning methods can sharpen searches for scalar extensions of the SM in the clean environment of a photon collider \cite{Feickert:2023, Bhattacherjee:2024, Soldner:2001, Berger:2025}.
\section{The Type-III Two-Higgs-Doublet Model}
The 2HDM-III extends the SM scalar sector by a second doublet and lets both doublets, $\Phi_1$ and $\Phi_2$, couple to every fermion type. The Yukawa sector is therefore the part of the Lagrangian of main interest here, and we summarize it below; the gauge-kinetic and scalar-potential terms are standard and are not repeated. The most general Yukawa interaction reads
\begin{equation}
-L_{Y}=\bar{Q_L}Y^u_1 U_R\tilde{\Phi_1} + \bar{Q_L}Y^u_2 U_R\tilde{\Phi_2} + \bar{Q_L}Y^d_1 D_R\Phi_1 + \bar{Q_L}Y^d_2 D_R\Phi_2 + \bar{L}Y^l_1 l_R\Phi_1 + \bar{L}Y^l_2 l_R\Phi_2 + H.C. 
\end{equation}
where $Q_{L}=(u_{L}, d_{L})$ and $L=(\ell_{L},\nu_{L})$ are $SU(2)_{L}$ doublets and $Y^{f,\ell}_{1,2}$ are $3\times 3$ Yukawa matrices. Because both doublets couple to the same fermion type, flavor-changing neutral currents (FCNCs) can arise at tree level; they must be kept small while the flavor-violating Higgs couplings of interest are retained \cite{atlas2012observation,collaboration2015precise,arhrib2017prospects,ross1975neutral,veltman1976second}. A frequently adopted remedy is a flavor symmetry acting on the Yukawa textures, which keeps the off-diagonal entries small and thus suppresses FCNCs, while still permitting flavor-dependent phenomena. In this scenario the non-diagonal couplings take the form
\begin{equation*}
\begin{aligned}
-L^{III}_Y ={}& \sum\limits_{f=u,d,l} \frac{m_j^{f}}{v}
\Big[(\xi^f_h)_{ij}\,\bar{f}_{Li}f_{Rj}\,h +
(\xi^f_H)_{ij}\,\bar{f}_{Li}f_{Rj}\,H - i(\xi^f_A)_{ij}\,\bar{f}_{Li}f_{Rj}\,A\Big] \\
&+\frac{\sqrt{2}}{v}\sum\limits_{k=1}^{3}\bar{u}_i\Big[m^u_i(\xi^{u*}_A)_{ki} V_{kj}P_L + V_{ik}(\xi^d_A)_{kj} m^d_j P_R\Big] d_j H^+
+\frac{\sqrt{2}}{v}\bar{\nu}_i(\xi^l_A)_{ij}m^l_jP_R\,l_j H^+ + \text{H.c.}
\end{aligned}
\end{equation*} 
The BPs are required to satisfy vacuum stability, unitarity and perturbativity. Flavor observables, notably $b\to s\gamma$ and $B\to\tau\nu$, set lower limits on $M_{H^{\pm}}$ and $\tan\beta$ and are taken into account as well \cite{HFLAV:2023, Enomoto:2016}. The mixing angle $\alpha$ and the ratio of vacuum expectation values $\tan\beta$ are fixed in the BPs such that the couplings of the $125~\text{GeV}$ Higgs boson agree with the LHC signal-strength measurements.

\section{Collider Phenomenology}
Benchmark points (BPs) provide concrete parameter choices for 2HDM predictions and for comparison with data. For the CP-conserving 2HDM considered here, the scalar parameters and $M_{H^{\pm}}$ are scanned over $100$--$190~\text{GeV}$, with the SM-like Higgs mass set to $125.09~\text{GeV}$. The BPs satisfy tree-level unitarity, perturbativity and vacuum stability, and are consistent with the electroweak oblique parameters. The BPs in Table~\ref{tab:placeholder} are adopted from previous studies of the 2HDM-III \cite{arroyo2025hunting, arroyo2025prospects} and have been re-checked against the theoretical and experimental constraints described above.

\begin{figure}[!ht]
    \centering
    \includegraphics[width=5cm,height=5cm]{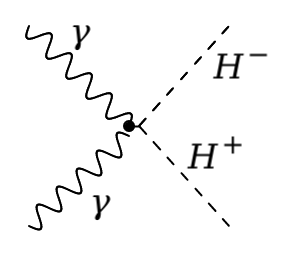}
        \includegraphics[width=5cm,height=5cm]{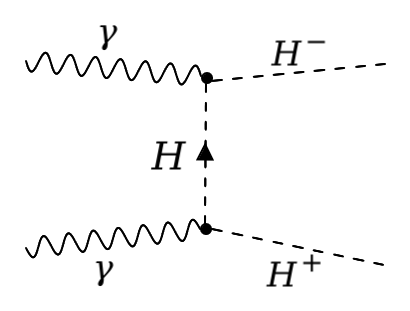}
    \caption{Tree-level Feynman diagrams for charged Higgs pair production through photon fusion at an $e^{+}e^{-}$ linear collider.}
    \label{feyn}
\end{figure}

\begin{figure}[t]
    \centering
    \includegraphics[width=8cm]{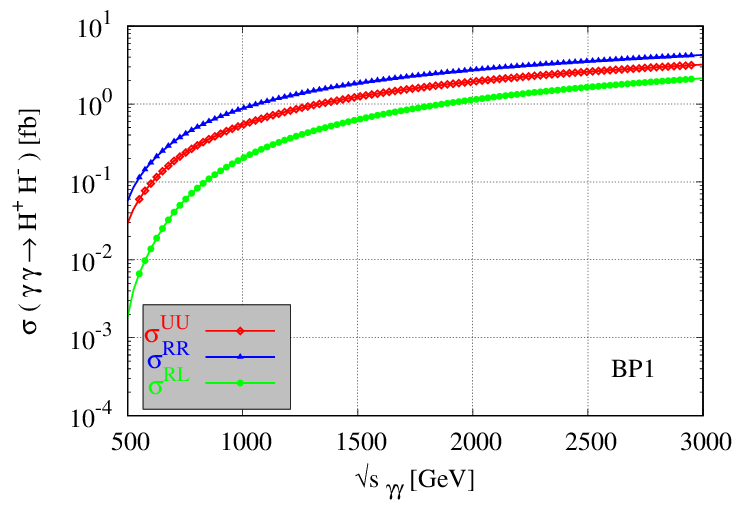}
        \includegraphics[width=8cm]{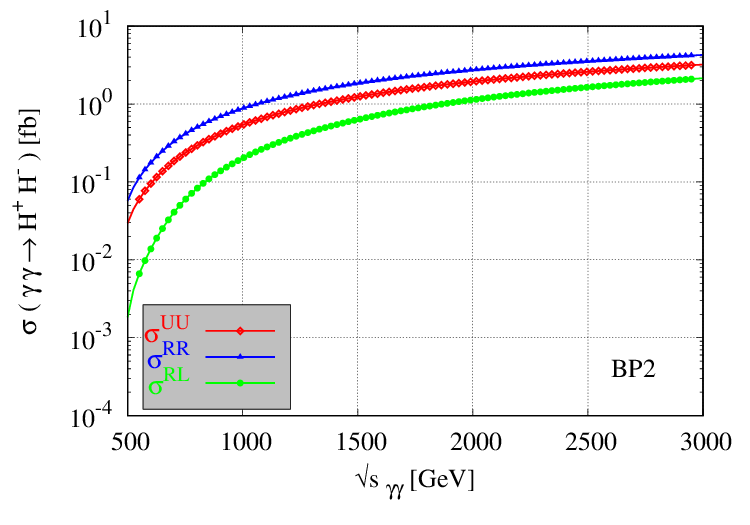}
    \caption{Production cross-section of BP1 (left) and BP2 (right) as a function of the center-of-mass energy.}
    \label{BP-CX}
\end{figure}

\begin{figure}[t]
    \centering
    \includegraphics[width=8cm]{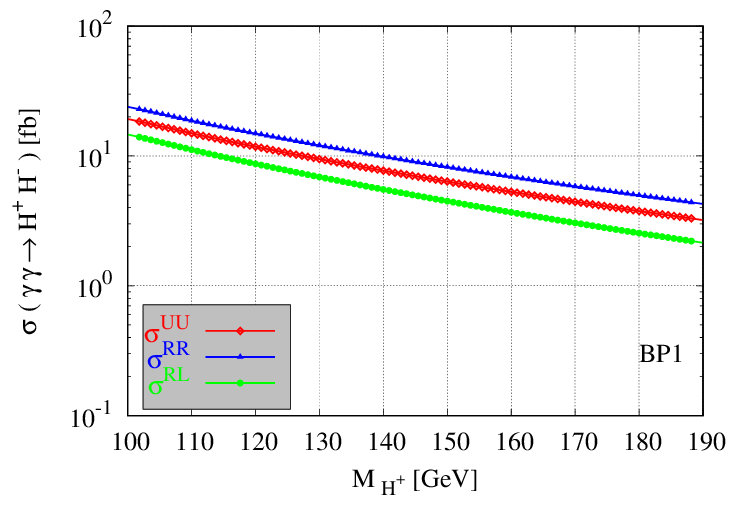}
        \includegraphics[width=8cm]{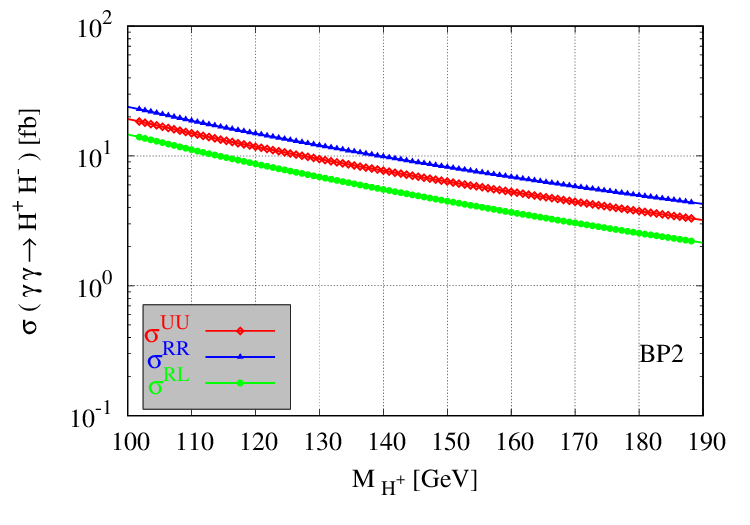}
    \caption{Production cross-section of BP1 (left) and BP2 (right) at $\sqrt{s}=3~\text{TeV}$ for right--right ($RR$), left--left ($LL$), right--left ($RL$) and unpolarized photon beams.}
    \label{BP-CX2}
\end{figure}

\begin{table}[h]
    \centering
    \begin{tabular}{p{90pt}  p{90pt} p{80pt} p{80pt} p{30pt}}\hline\hline
         \textbf{Benchmarks}&$\mathbf{M_{H^{\pm}}(GeV)}$&$\mathbf{\cbma}$&$\mathbf{m^{2}_{12}}$&$\mathbf{\tb}$ \\\hline
        \textbf{BP1} & 100\dots 190& -0.2&24752.48&10\\\hline
        \textbf{BP2} & 100\dots 190& -0.1&17955.1122&20\\\hline\hline
    \end{tabular}
    \caption{Benchmark scenarios for the study of the charged Higgs boson \cite{arroyo2025hunting, arroyo2025prospects}. For BP1, $M_{A}=300~\text{GeV}$ and $M_{H^{0}}=500~\text{GeV}$; for BP2, $M_{A}=400~\text{GeV}$ and $M_{H^{0}}=600~\text{GeV}$.}
    \label{tab:placeholder}
\end{table}
We set $\chi_{\tau}=1$ for the lepton-Yukawa sector in all BPs \cite{arroyo2025hunting, arroyo2025prospects}. The Feynman diagrams for $H^{+}H^{-}$ production are shown in Fig.~\ref{feyn}. MadGraph5 v3.4.2 \cite{alwall2011madgraph} is used to simulate the signal and the backgrounds across a mass scan over $M_{H^{\pm}}\in[100,190]~\text{GeV}$, with $1.0\times10^{5}$ events per process to limit statistical fluctuations. The main irreducible backgrounds are $\gamma\gamma\to WWZ$, $\gamma\gamma\to t\bar{t}Z$ and $\gamma\gamma\to t\bar{t}$. The $\tau\nu$ and $tb$ final states are selected because they are regarded as the leading modes over $100$--$190~\text{GeV}$ in the 2HDM-III. The sensitivity of photon colliders to Higgs-pair production in other versions of the 2HDM has also been explored \cite{Arhrib:2025}. Figure~\ref{BP-CX} gives the cross-section for BP1 and BP2, which rises with increasing energy.
Cross-sections are evaluated at $\sqrt{s}=3~\text{TeV}$ for photons radiated elastically from the lepton beams of a linear collider, with right-handed, left-handed and unpolarized photon beams. In Fig.~\ref{BP-CX2}, $RR$, $LL$ and $RL$ stand for the right--right ($++$), left--left ($--$) and right--left ($+-$) beam configurations. Branching ratios and decay widths are obtained with \textbf{2HDMC 1.8.0} (Two-Higgs-Doublet Model Calculator) \cite{rathsman20112hdmc}, and the plots are produced with Gnuplot \cite{williamsgnuplot}. In BP1, $\mathrm{BR}(H^{+}\to\tau\nu_{\tau})$ falls from $4.62067338\times10^{-3}$ at $M_{H^{\pm}}=100~\text{GeV}$ to $1.79110173\times10^{-4}$ at $M_{H^{\pm}}=190~\text{GeV}$. The corresponding values (from $100$ to $190~\text{GeV}$) are $8.16184540\times10^{-1}\to2.81231614\times10^{-2}$ for $b\bar{c}$, $1.64394263\times10^{-1}\to5.64603515\times10^{-3}$ for $c\bar{s}$, and $1.63489443\times10^{-5}\to6.33442349\times10^{-7}$ for $\mu\nu_\mu$. In BP2, $\mathrm{BR}(H^{+}\to\tau\nu_{\tau})$ decreases from $2.90635149\times10^{-4}$ at $M_{H^{\pm}}=100~\text{GeV}$ to $1.15557745\times10^{-5}$ at $M_{H^{\pm}}=190~\text{GeV}$. We consider the pairs $H^{\pm}H^{\mp}$ from photon fusion whose decays yield charged leptons together with missing transverse energy ($\cancel{E}_{T}$) from the undetected neutrinos.

\begin{figure}[t]
    \centering
    \includegraphics[width=8cm]{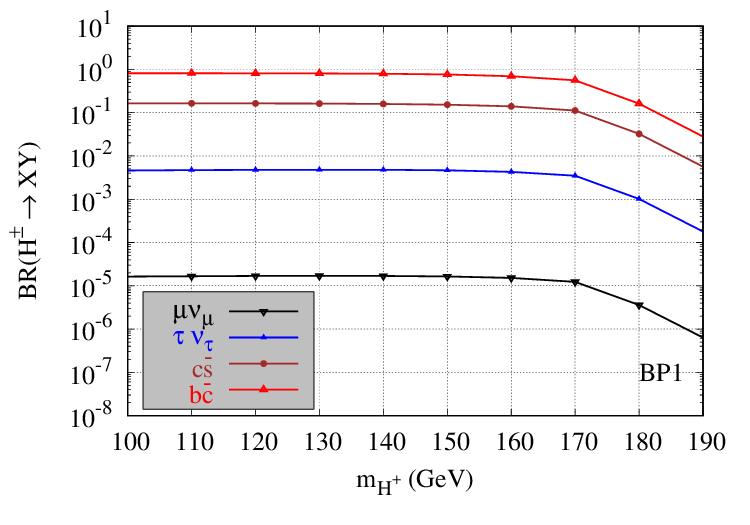}
    \includegraphics[width=8cm]{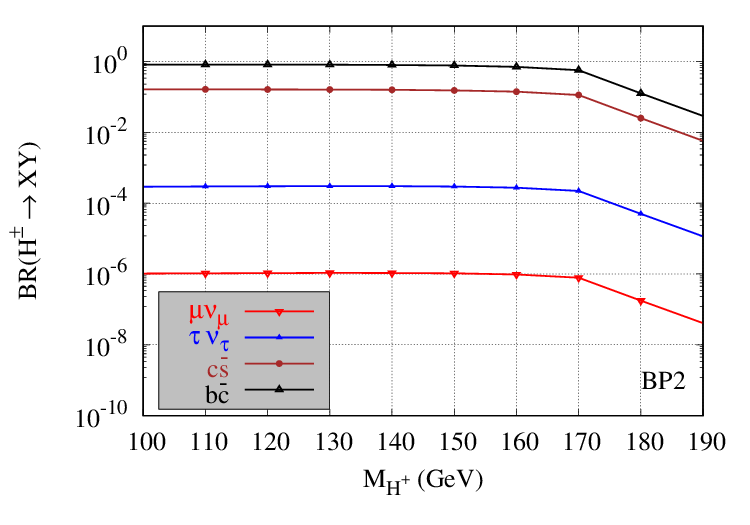}
   \caption{Branching ratios of the charged Higgs boson for the benchmark points.}
   \label{BP-BR}
\end{figure}
\begin{figure}[t]
   \centering
    \includegraphics[width=8cm]{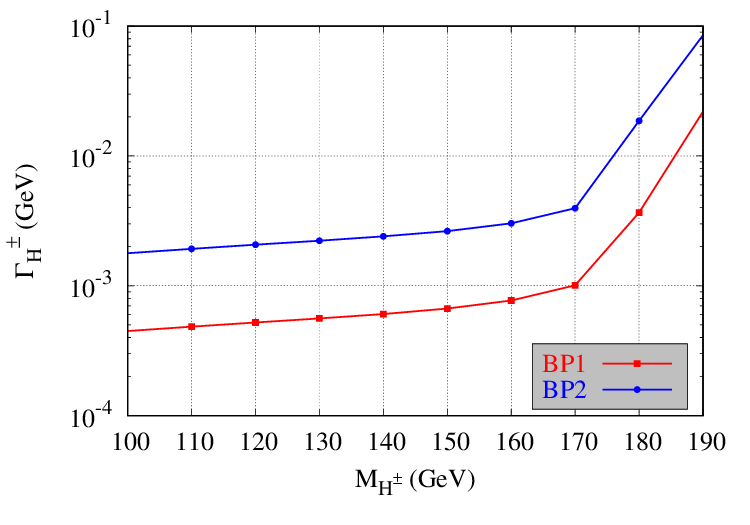}
    \includegraphics[width=8cm]{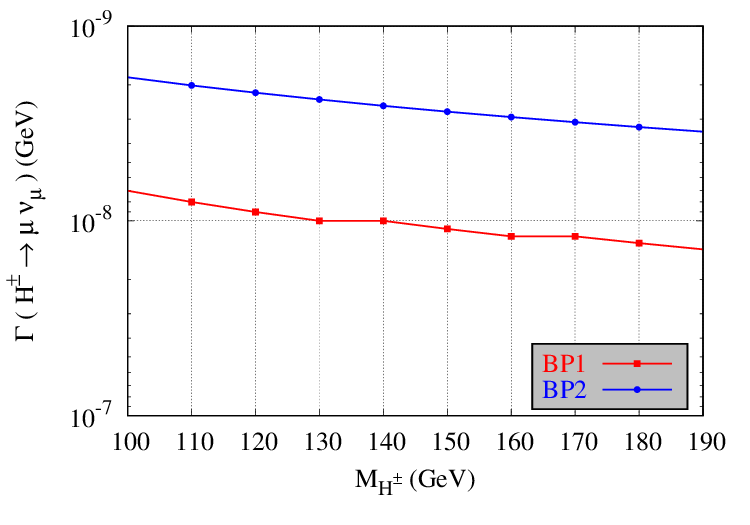}
    \caption{Total and partial widths for $H^{\pm}$ decays in the benchmark points.}
   \label{decay}
\end{figure}
\begin{figure}[t]
    \centering   
       \includegraphics[width=5cm,height=5cm]{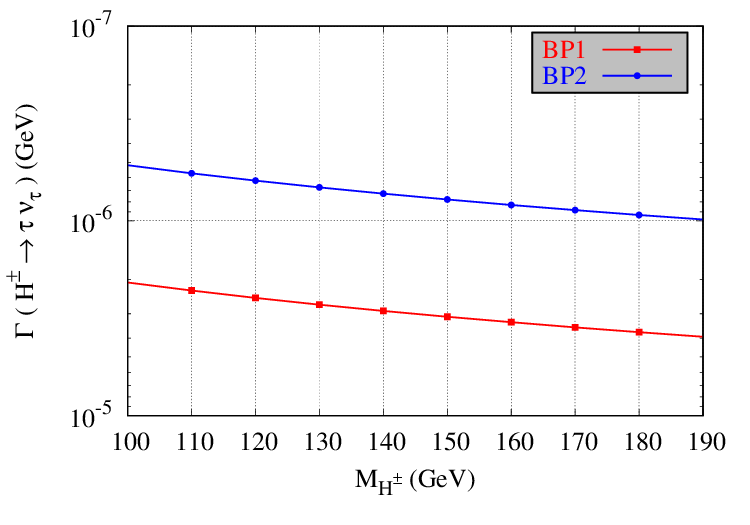}
   \includegraphics[width=5cm,height=5cm]{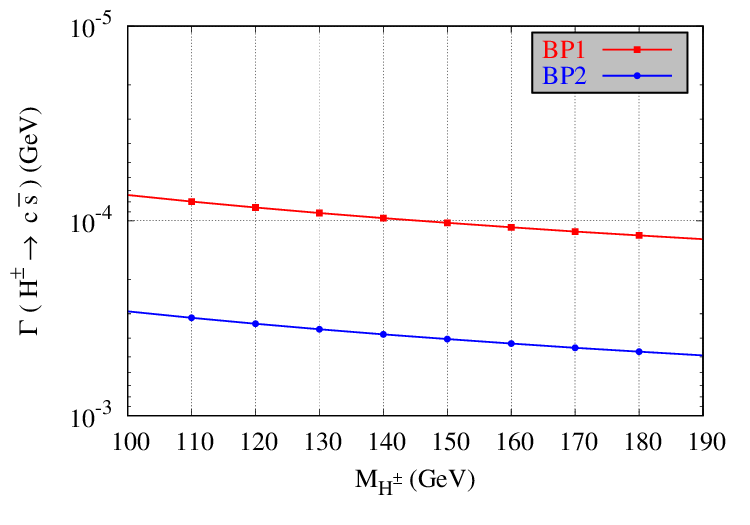}
   \includegraphics[width=5cm,height=5cm]{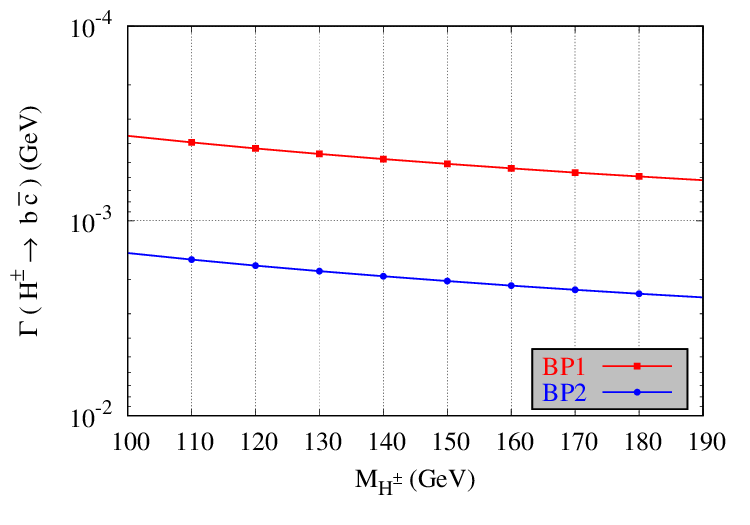}
   \caption{Partial widths of the $H^{\pm}$ decay modes in the benchmark points.}
   \label{partialdecay}
\end{figure}
\begin{figure}[t]
    \centering
    \includegraphics[width=8cm]{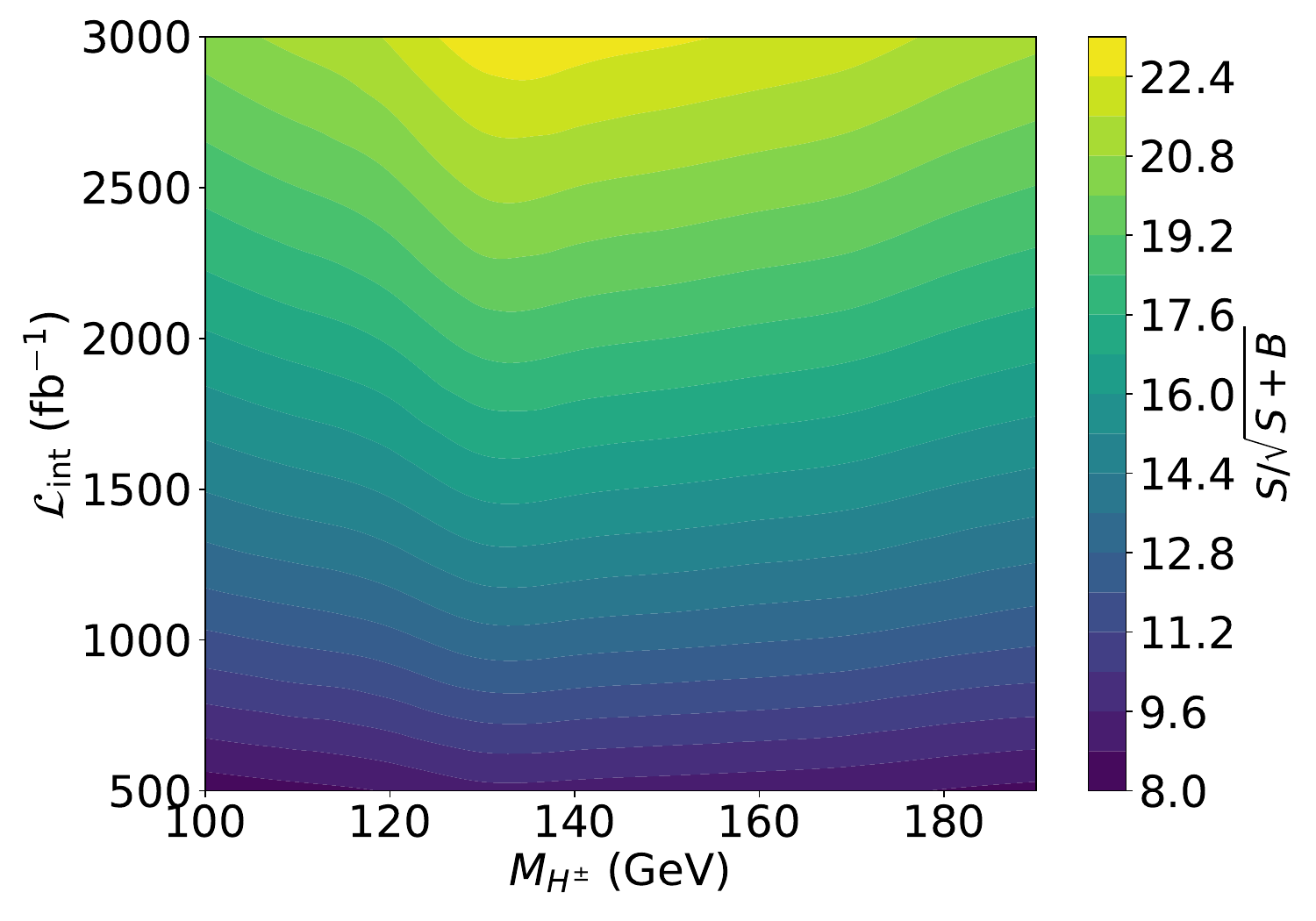}
        \includegraphics[width=8cm]{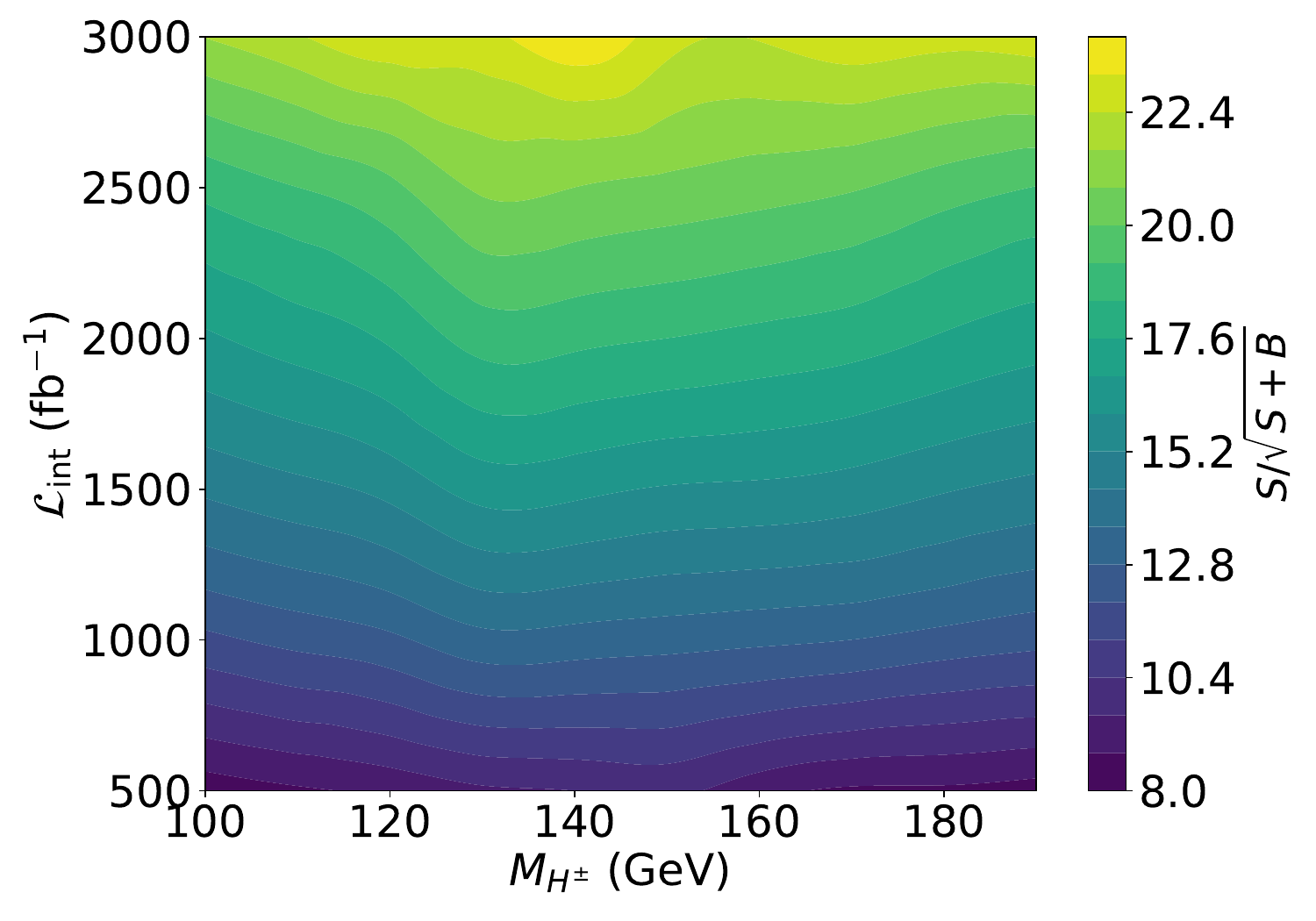}
    \caption{Significance for BP1 (left) and BP2 (right) as a function of the charged Higgs mass and the integrated luminosity.}
    \label{significance}
\end{figure}
The starting point is the list of kinematically allowed $H^{\pm}$ decay channels in both BPs. The total width is obtained by adding all partial widths,
\begin{equation}
    \Gamma=\sum_{j}\Gamma_{j},
\end{equation}
where, in natural units, $\Gamma$ is measured in GeV and the lifetime is $\tau=1/\Gamma$. Figure~\ref{decay} shows that $\Gamma_{H^{\pm}}$ grows slowly with $M_{H^{\pm}}$ as further decay channels open up. BP2 gives a somewhat larger width, which suggests stronger couplings or heavier fermions in the final state. With values between $10^{-4}~\text{GeV}$ and $10^{-1}~\text{GeV}$, the width is small, i.e. $H^{\pm}$ is a narrow state throughout the mass interval studied. The same figure shows that the purely leptonic partial width is tiny for all BPs, between $10^{-9}~\text{GeV}$ and $10^{-8}~\text{GeV}$, in line with the small muon Yukawa coupling; its irregular mass dependence is attributed to phase-space effects.

According to Fig.~\ref{partialdecay}, $H^{\pm}\rightarrow\tau\nu_{\tau}$ is the dominant leptonic channel among the fermionic decays, its size being set by the $\tau$ Yukawa coupling and a partial width between $10^{-7}~\text{GeV}$ and $10^{-5}~\text{GeV}$, depending on the BP. Fermionic channels thus dominate at low charged Higgs masses, as expected in the 2HDM, which supports the choice of BPs. The significance grows with both $M_{H^{\pm}}$ and the integrated luminosity (Fig.~\ref{significance}).

\section{Signal and Background Processes}
The signal is the photon-fusion process
\begin{equation}
    \gamma~\gamma\rightarrow H^{+}~H^{-}\nonumber
\end{equation}
and we use the $tb$ and $\tau\nu_{\tau}$ decays of the charged Higgs boson,
\begin{eqnarray}
    H^{\pm}\rightarrow t~b, ~~~~ t\rightarrow W~b,~~~ W\rightarrow \tau~\nu_{\tau}\nonumber
\end{eqnarray}
The final state then contains two tau leptons, several jets (including $b$-jets) and large missing transverse energy ($\cancel{E}_{T}$) from the neutrinos, a distinctive signature for $M_{H^{\pm}}=190~\text{GeV}$.

Backgrounds originate from SM electroweak and top-quark production in photon collisions:
\begin{eqnarray}
    \gamma~\gamma\rightarrow W~W~Z~~~,~~ W\rightarrow \tau~\nu_{\tau}\nonumber
\end{eqnarray}
\begin{eqnarray}
    \gamma~\gamma\rightarrow t~\bar{t}~Z~~~,~~ t\rightarrow b~\tau~\nu_{\tau}\nonumber
\end{eqnarray}
\begin{eqnarray}
    \gamma~\gamma\rightarrow t~\bar{t}~~~,~~ t\rightarrow b~\tau~\nu_{\tau}\nonumber
\end{eqnarray}
In these processes the $Z$ boson decays hadronically,
\begin{eqnarray}
    Z\rightarrow j~j\nonumber
\end{eqnarray}
and, since their final states mimic that of the signal, the backgrounds are irreducible.

\section{Multivariate Analysis of Charged Higgs Boson Production}
Machine learning is now a standard tool in searches for new physics at high-energy colliders \cite{Thais:2022, Duarte:2018}. We use the TMVA package \cite{TMVA, TMVA4}, a multivariate-analysis toolkit distributed with ROOT \cite{brun1997root}, which provides a common interface for training, testing and comparing many classifiers on labeled signal and background samples. Multivariate techniques have recently shown considerable promise for probing heavy charged Higgs bosons at $\gamma\gamma$ colliders \cite{Ahmed:2025}.

\subsection{Input Variables and Kinematical Cuts}

Table~\ref{inputvariables} lists the kinematic variables supplied to the classifiers.
\begin{table}[h]
\renewcommand{\arraystretch}{1.2}
\caption{Observables supplied as inputs to the classifiers.}
\centering
\begin{tabular}{p{4cm} p{8cm}} \hline\hline
\textbf{Variables} & \textbf{Name} \\ \hline

$\mathbf{P_{T}^{\text{jet}}}$ & Transverse momentum \\ \hline
$\mathbf{\eta_{\text{jet}}}$ & Pseudorapidity \\ \hline
$\mathbf{N_{\text{jet}}}$ & Number of jets \\ \hline
$\mathbf{\cancel{E}_T}$ & Missing transverse energy \\ \hline
$\mathbf{H_T}$ & Scalar sum of $P_T$ \\ \hline
$\mathbf{M_{\text{jet}}}$ & Jet mass \\ \hline
$\mathbf{\Delta \eta_{\text{jet}}}$ & Pseudorapidity separation \\ \hline
$\mathbf{\Delta \phi_{\text{jet}}}$ & Azimuthal angle separation \\ \hline

\end{tabular}
\label{inputvariables}
\end{table}
The preselection cuts are given in Table~\ref{inputvariablescuts}. The sample is divided into two equal halves, $50\%$ for training and $50\%$ for testing, so that overtraining can be monitored. All classifiers are trained with TMVA \cite{TMVA, TMVA4} inside ROOT. In the variable ranking, $\cancel{E}_T$ and $H_T$ emerge as the most discriminating inputs for both the BDT and the MLP.
\begin{table}[h]
\def\arraystretch{1}
 \caption{Preselection cuts applied to the classifier inputs.}
    \centering
    \begin{tabular}{p{80pt}|| p{80pt} p{80pt} p{80pt} p{80pt}}\hline\hline
        \textbf{Variables}&~~~~~~~~$\mathbf{P_{T}^{jet}}$&~~~~~~~~$\mathbf{\eta_{jet}}$ &~~~~~~~~~~$\mathbf{N_{jet}}$&~~~~~~~~~~$\mathbf{\cancel{E}_{T}}$\\\hline  
     \textbf{Cuts }&~~~~~~$>20~GeV$&~~~~~~$<2.6$&~~~~~~~~~~$\geq 3$&~~~~~~$>50~GeV$\\\hline
    \end{tabular}
      \label{inputvariablescuts}
\end{table}
\subsection{Boosted Decision Tree (BDT)}
Four classifiers are compared: BDT, MLP, Likelihood and LikelihoodD. A decision tree applies a chain of binary cuts, each on a single variable. Starting from the root node, every node uses the cut that best separates signal from background, and a node is no longer split once it would fall below the minimum node size (\texttt{MinNodeSize}). Each terminal node (leaf) is labeled according to its purity $p$, the fraction of signal events it contains: $p>0.5$ gives signal ($+1$) and $p<0.5$ gives background ($-1$) \cite{coadou2022boosted}. A single tree is a weak classifier, so many trees are combined by boosting. In AdaBoost, events misclassified by one tree receive larger weights when the next tree is trained, and the BDT output is the weighted combination of all trees. An event is called signal if its BDT output exceeds a threshold $y_{cut}$; scanning $y_{cut}$ yields pairs of signal efficiency $\epsilon_{sig,eff}$ and background rejection $1-\epsilon_{bkg,eff}$ \cite{speckmayer2010toolkit}.

\subsection{Likelihood Ratio}
The projective likelihood method estimates, from the training sample, a signal and a background probability density function (PDF) for each input variable, $P_{S,i}$ and $P_{B,i}$, each normalized to unity,
\begin{equation}
    \int^{\infty}_{-\infty}P_{S/B,i}(x_{i})dx_{i}=1.
\end{equation}
For event $j$, the PDFs are multiplied over the $n_{var}$ variables,
\begin{equation}
    L_{S/B}(j)=\prod_{i=1}^{n_{var}}P_{S/B,i}(x_{i}(j)),
\end{equation}
and the classifier output is the ratio
\begin{equation}
    y_{L}(j) = \dfrac{L_{S}(j)}{L_{S}(j)+L_{B}(j)}.
\end{equation}
Since the product treats the variables as independent, correlations between them are ignored, which degrades the performance when such correlations are present.
\subsection{Likelihood with Decorrelation (LikelihoodD)}
Linear correlations among the inputs can be reduced before the PDFs are built. From the covariance matrix $C$ of the training sample, TMVA constructs its square root $C^{'}$, with $C=(C^{'})^{2}$, by diagonalizing $C$,
\begin{equation}
    D=S^{T}CS\impliedby C^{'}=S\sqrt{D}S^{T},
\end{equation}
where $D$ is a diagonal matrix and $S$ is a symmetric matrix. Applying the inverse of $C^{'}$ to the vector of input variables $\texttt{x}$ removes the linear correlations,
\begin{equation}
    \texttt{x} \longmapsto (C^{'})^{-1} \texttt{x}.
\end{equation}
The removal is complete only for Gaussian-distributed, linearly correlated variables; any remaining nonlinear dependence is not corrected. The hyperparameters used in the analysis are listed in Table~\ref{hyperparameters}.
\begin{table}[h]
\def\arraystretch{1}
    \centering
    \begin{tabular}{p{90pt}|| p{80pt} p{80pt} p{80pt} p{80pt}}\hline\hline
        \textbf{Hyperparameters}&\textbf{NTrees}&\textbf{MaxDepth} &\textbf{MinNodeSize}&\textbf{AdaBoostBeta}\\\hline  
     \textbf{Values}&~~~~800&~~~~~5&~~~~~$2.5\%$&~~~~~~0.8\\\hline
    \end{tabular}
   \caption{Hyperparameters adopted in the multivariate analysis.}
    \label{hyperparameters}
\end{table}
\subsection{Multilayer Perceptron (MLP)}
The MLP is a feed-forward neural network whose neurons are arranged in layers: the first layer holds one neuron per input variable ($n_{var}$ in total) together with a bias neuron; it is followed by one or more hidden layers, whose sizes are set by the \texttt{HiddenLayers} option and which also contain a bias node, and by a final layer that provides the response $y_{ANN}$. Each link joining two neurons has its own weight. Allowing connections only between successive layers keeps the number of weights manageable, whereas a fully connected set of $n$ neurons could have up to $n^{2}$ directed links \cite{TMVA}. Each neuron first sums its weighted inputs through a synapse function $\kappa$ and then applies an activation function $\alpha$, so that its response is $\rho=\alpha\circ\kappa$. In the simplest case of a single hidden layer with $\tanh$ activation and without bias nodes, the network output is
\begin{equation}
    y_{ANN}=\tanh\biggl(\sum^{j=1}_{n}y^{(2)}_{j}\omega^{(2)}_{i,1}\biggr)=\tanh\biggl[\sum^{n_{h}}_{j=1}\tanh\biggl(\sum^{n_{var}}_{j}x_{i}\omega^{(1)}_{i,j}\omega^{(2)}_{i,1}\biggr)\biggr] \label{yann}
\end{equation}
with $n_{h}$ hidden-layer nodes and $n_{var}$ input variables.
\begin{table}[t]
\def\arraystretch{1.5}
    \centering
    \begin{tabular}{p{100pt} p{100pt} c}\hline\hline
        \textbf{MVA Classifier} & \textbf{AUC (with cut)}& \textbf{AUC (without cut)}\\\hline
         \textbf{MLP}& ~~~~~~~0.934 & 0.932\\\hline
         \textbf{BDT}&~~~~~~~0.933&0.930\\\hline
         \textbf{LikelihoodD}&~~~~~~~0.870& 0.871\\\hline
         \textbf{Likelihood}&~~~~~~~0.815& 0.814\\\hline
    \end{tabular}
    \caption{AUC values of the four classifiers, with and without the kinematic cuts.}
    \label{bkg_rej_sig_eff_table}
\end{table}
Training seeks the weights that minimize the error function $E$. Starting from random weights, they are updated iteratively against the gradient of $E$,
\begin{equation}
    \Vec{\omega}^{\rho+1}=\Vec{\omega}^{\rho}-\eta\nabla_{\Vec{\omega}}E,
\end{equation}
where the learning rate $\eta>0$ (the \texttt{LearningRate} option) sets the step size.
\begin{figure}[t]
\centering     
\subfigure[]{\includegraphics[width=80mm,height=60mm]{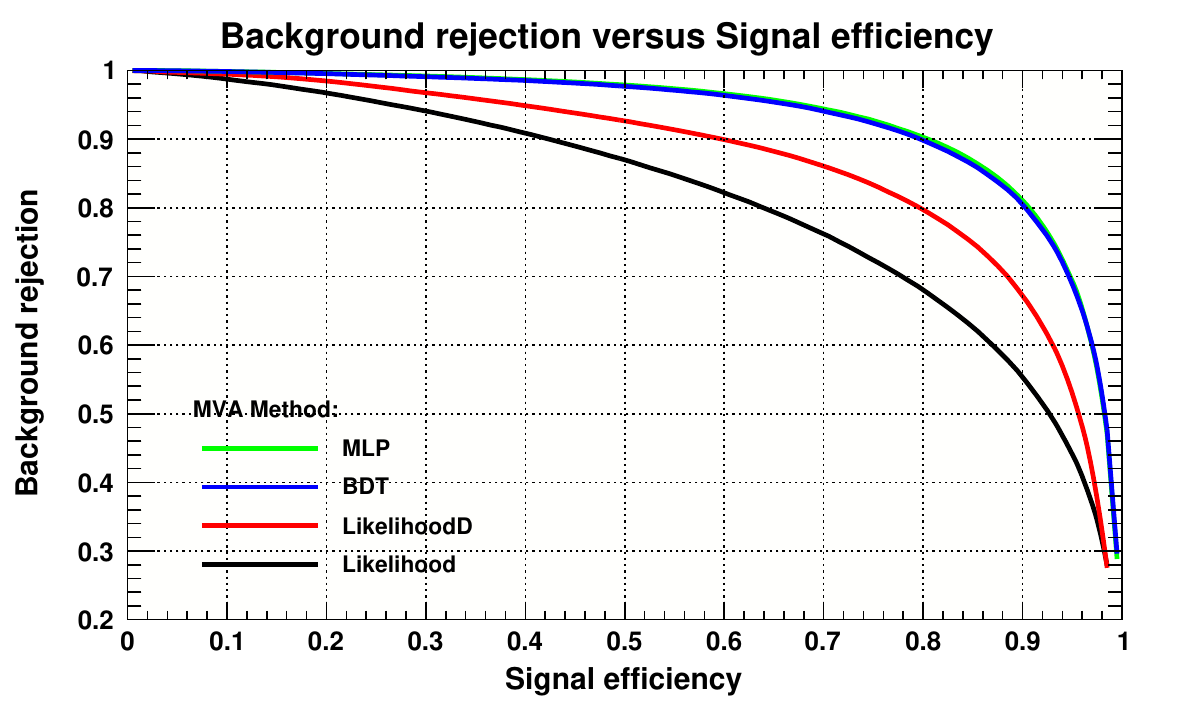}}
\subfigure[]{\includegraphics[width=80mm, height=60mm]{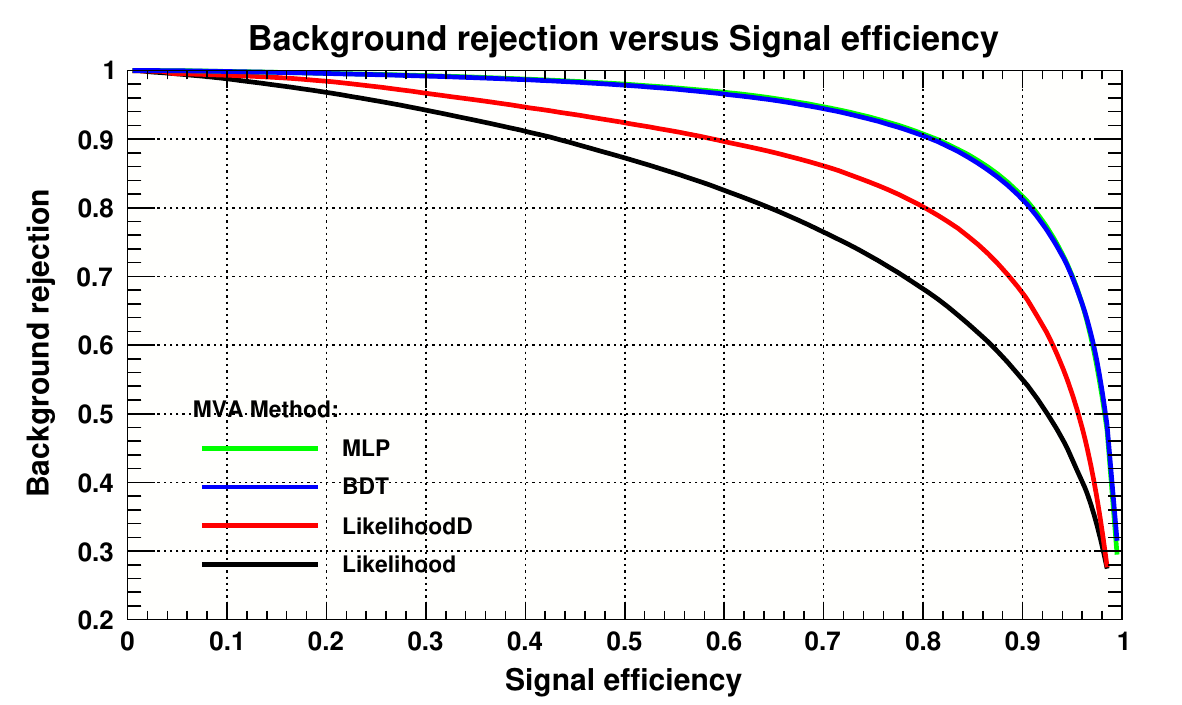}}
\caption{Background rejection versus signal efficiency for the four classifiers: (a) without cuts, (b) with cuts.}
\label{cutsifeffbkgrej}
\end{figure}
For the output layer, the updates follow from Eq.~(\ref{yann}),
\begin{equation}
    \Delta\omega^{(2)}_{i,1}=-\eta\dfrac{\partial E_{a}}{\partial\omega^{(2)}_{i,1}}=-\eta\dfrac{1}{2}\dfrac{\partial[(y_{ANN,a}-\hat{y}_{a})^{2}]}{\partial\omega^{(2)}_{i,1}}
\end{equation}
\begin{equation}
    \Delta\omega^{(2)}_{i,j}=-\eta(y_{ANN,a}-\hat{y}_{a})y_{ANN,a}(1-y_{ANN,a})y^{(2)}_{i,a}(1-y^{(2)}_{i,a})\omega^{(2)}_{i,1}\, x_{i}
\end{equation}
The procedure is repeated for every training event. After the learning phase, assuming that the weights minimizing $E$ have been found, the weights obtained from the last event define the classifier response through the neuron response function.
\begin{figure}[t]
\centering     
\subfigure[]{\includegraphics[width=80mm,height=60mm]{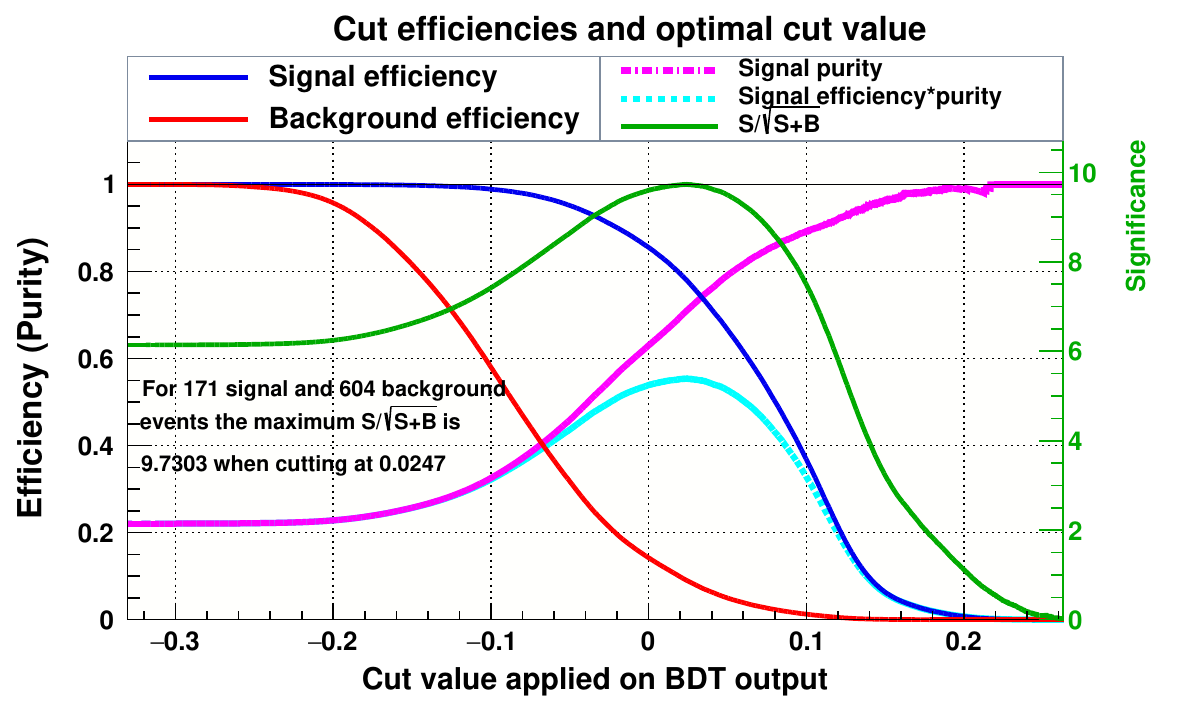}}
\subfigure[]{\includegraphics[width=80mm, height=60mm]{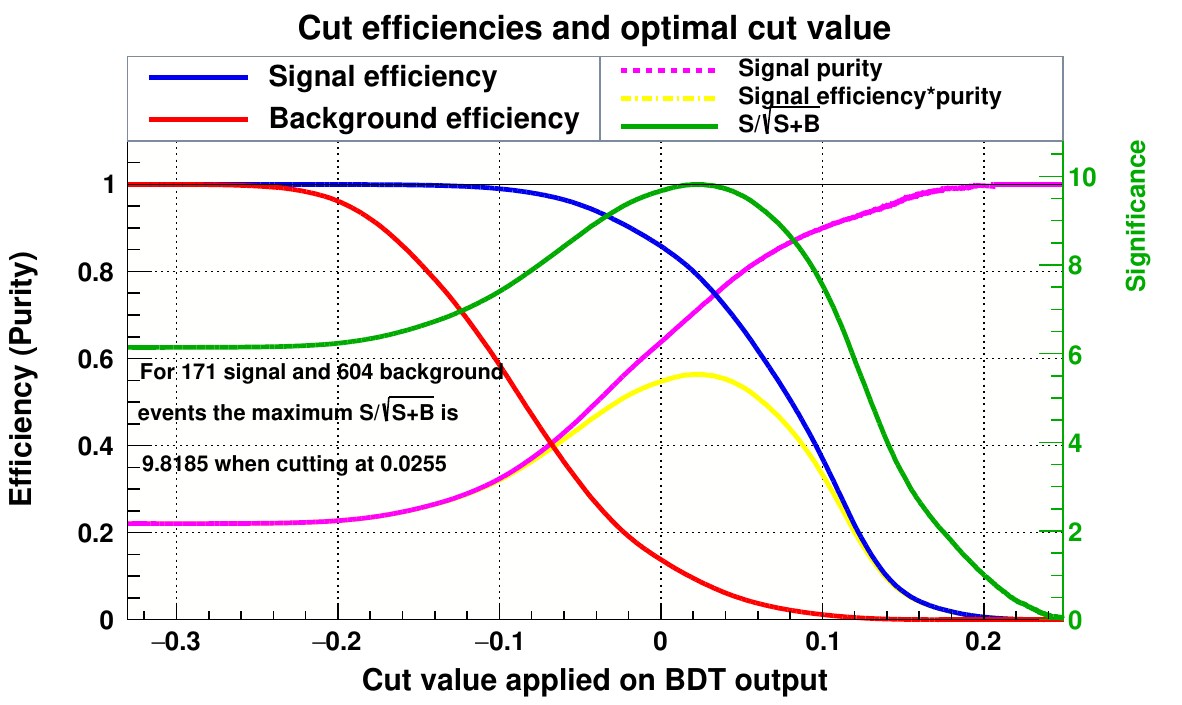}}
\caption{Cut efficiencies and signal significance for the BDT output: (a) without cuts, (b) with cuts.}
\label{cuteffbdt}
\end{figure}
\begin{figure}[t]
\centering     
\subfigure[]{\includegraphics[width=80mm,height=60mm]{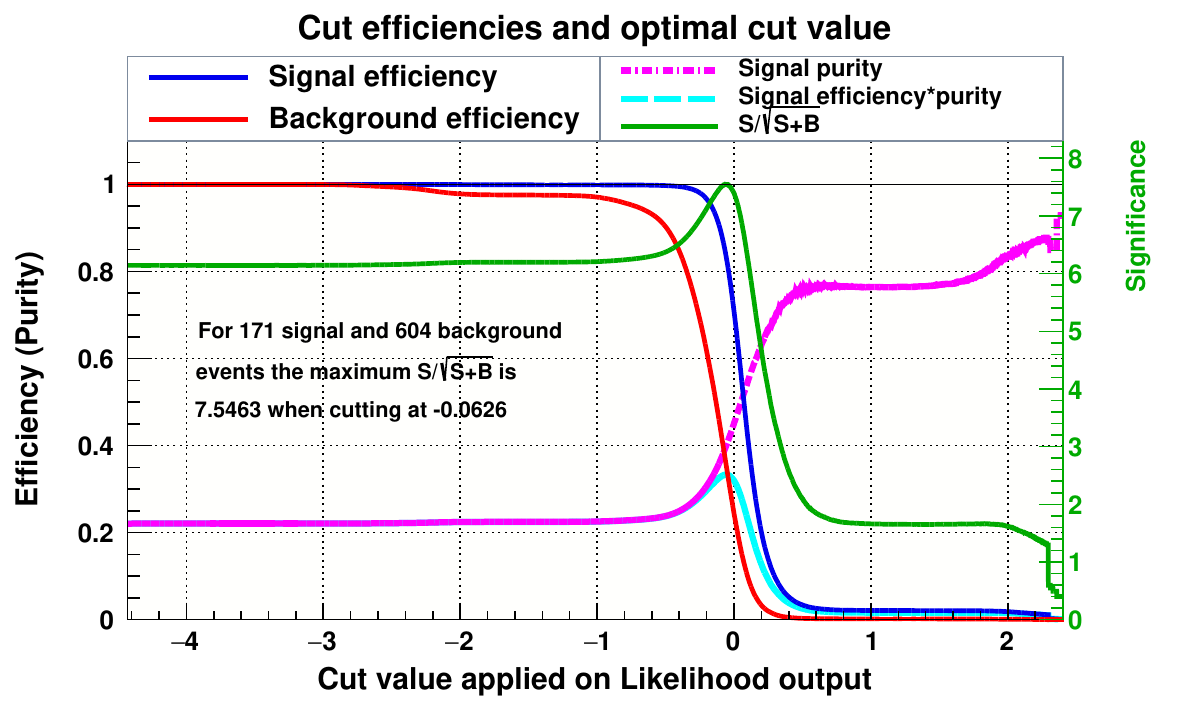}}
\subfigure[]{\includegraphics[width=80mm, height=60mm]{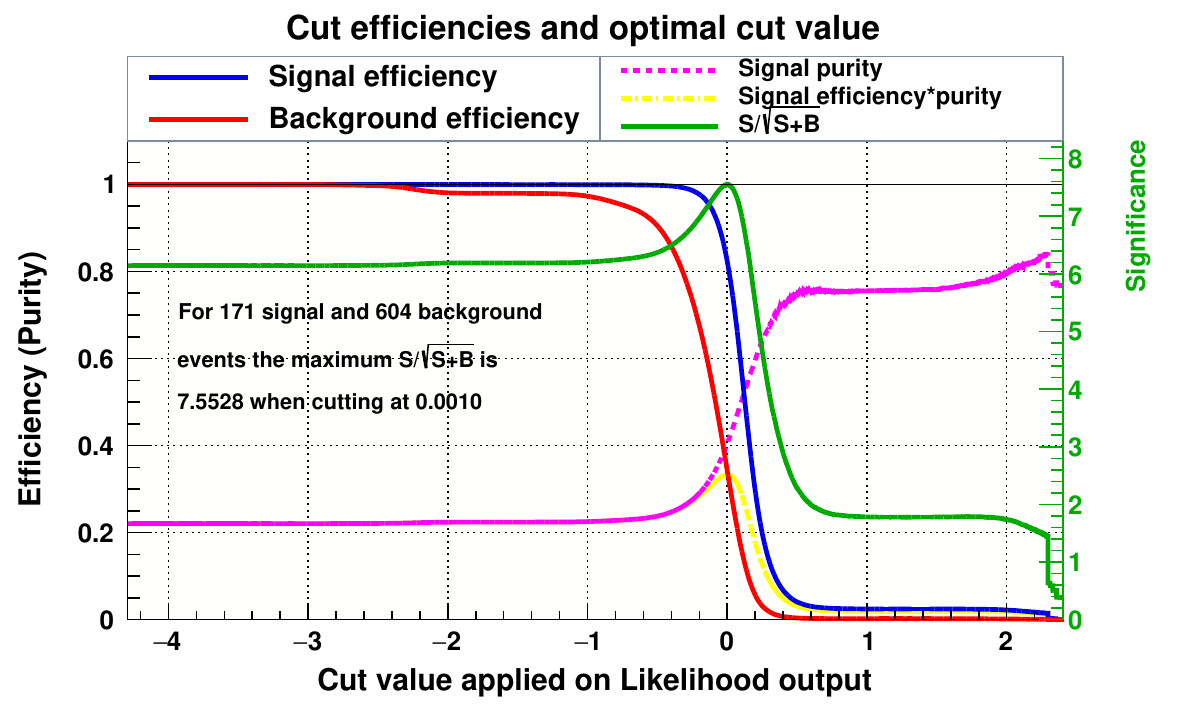}}
\caption{Cut efficiencies and signal significance for the Likelihood output: (a) without cuts, (b) with cuts.}
\label{cutefflikelihood}
\end{figure}
\begin{figure}[t]
\centering     
\subfigure[]{\includegraphics[width=80mm,height=60mm]{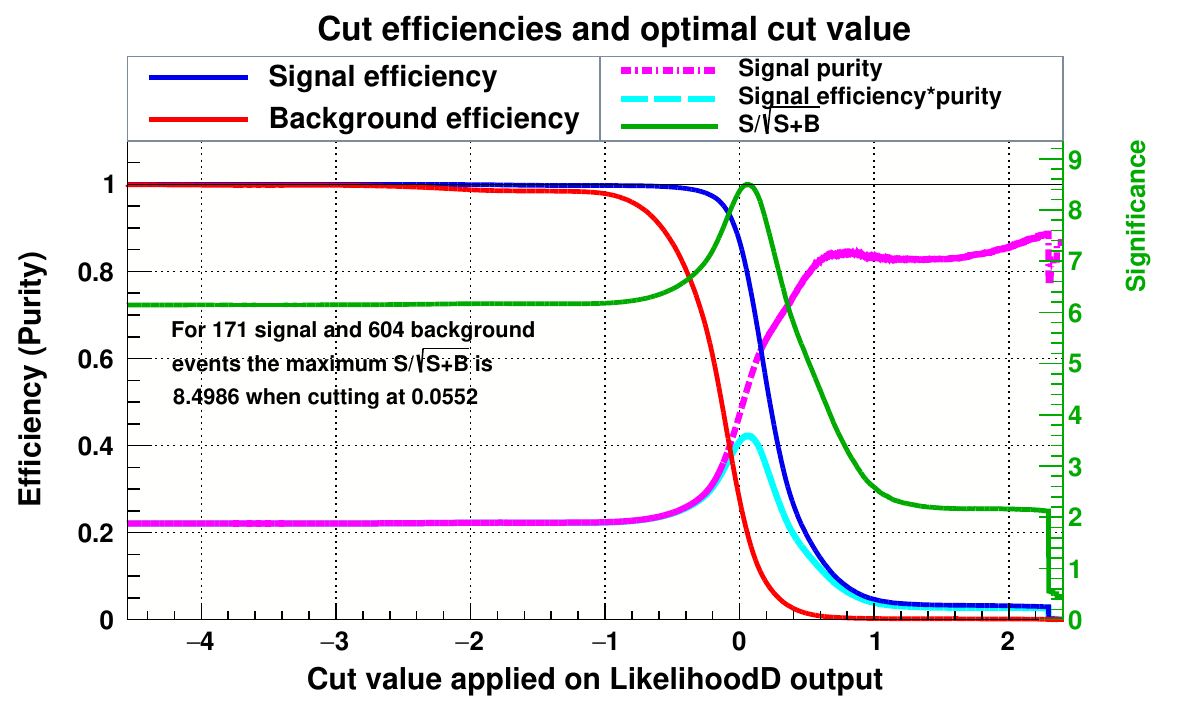}}
\subfigure[]{\includegraphics[width=80mm, height=60mm]{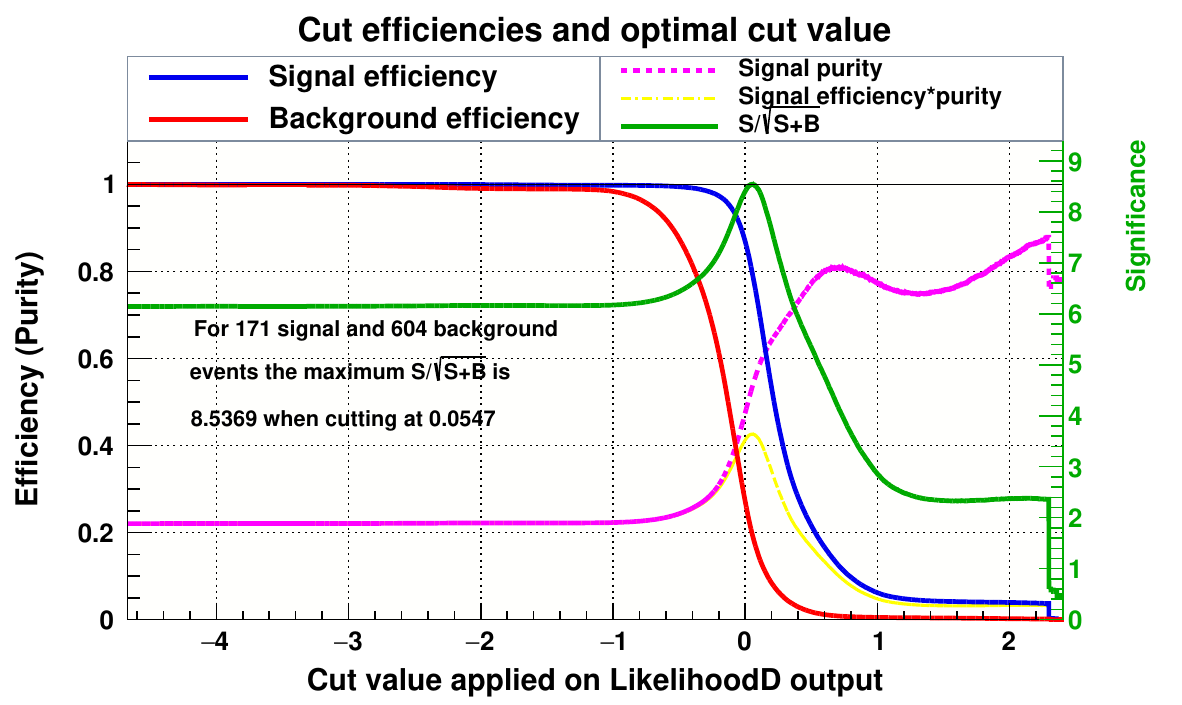}}
\caption{Cut efficiencies and signal significance for the LikelihoodD output: (a) without cuts, (b) with cuts.}
\label{cutefflikelihoodd}
\end{figure}
\begin{figure}[t]
\centering     
\subfigure[]{\includegraphics[width=80mm,height=60mm]{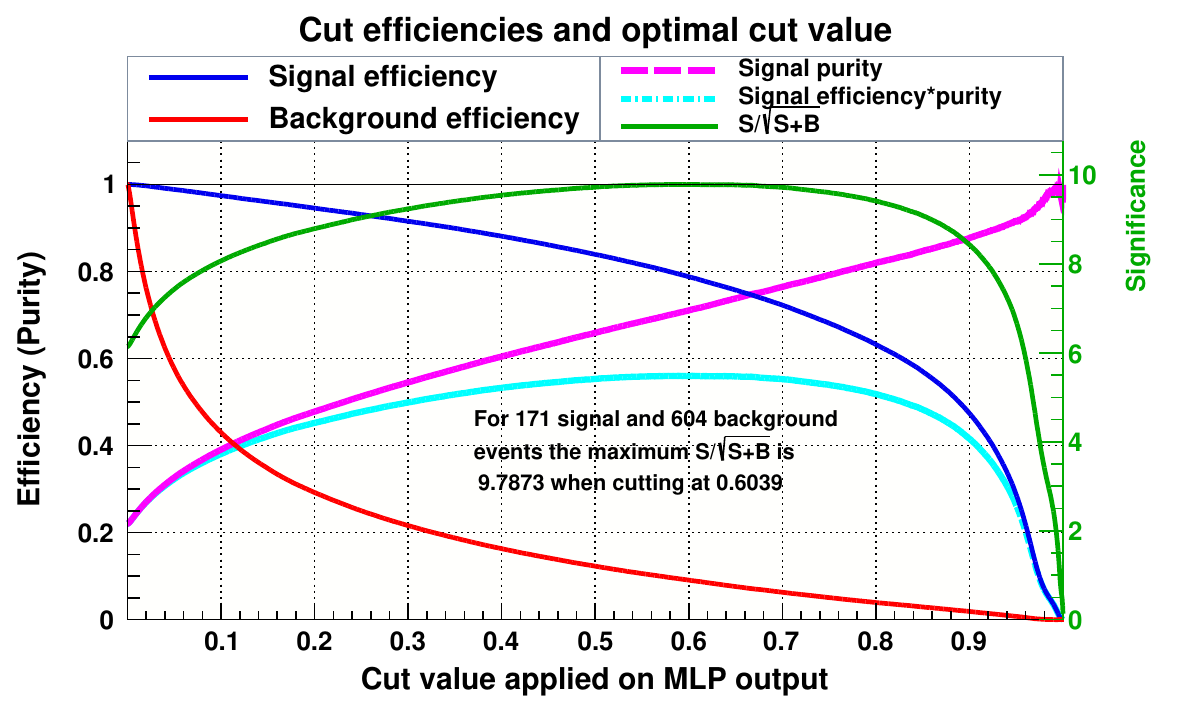}}
\subfigure[]{\includegraphics[width=80mm, height=60mm]{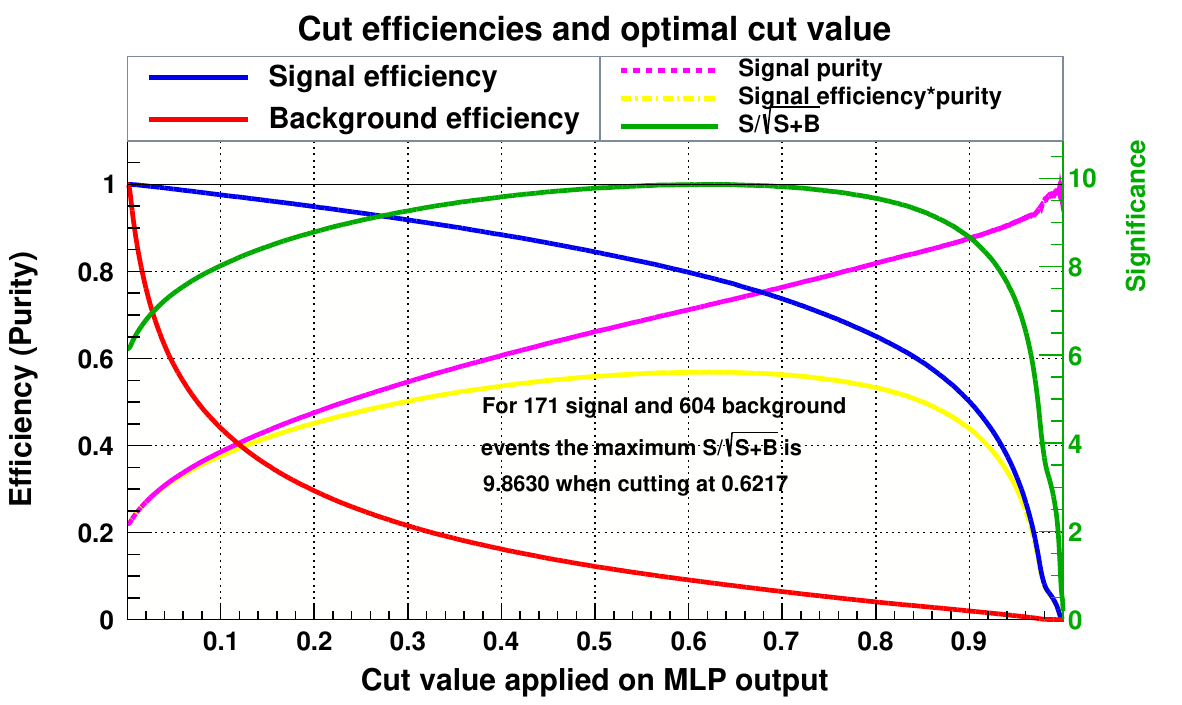}}
\caption{Cut efficiencies and signal significance for the MLP output: (a) without cuts, (b) with cuts.}
\label{cuteffmlp}
\end{figure}
\begin{table}[t]
\def\arraystretch{1.5}
    \centering
    \begin{tabular}{p{100pt} p{130pt} p{100pt} p{100pt}}\hline\hline
      \textbf{MVA  Classifier }&\textbf{Signal Significance}\newline\textbf{(with cuts)}&\textbf{Signal Significance}\newline\textbf{(without cuts)}\\\hline
        \textbf{MLP}&~~~~9.8630 &~~~~9.78728 \\\hline 
\textbf{ LikelihoodD}&~~~~8.5369&~~~~8.49856\\\hline\textbf{ Likelihood}&~~~~7.5528&~~~~7.54631\\\hline
\textbf{ BDT}&~~~~9.8185 &~~~~9.73028 \\\hline

    \end{tabular}
   \caption{Maximum signal significance reached by each classifier, with and without the cuts, for $\mathcal{L}_{int}=3000~\text{fb}^{-1}$.}
    \label{signalsig}
\end{table}
The samples correspond to 171 signal and 604 background events at $3000~\text{fb}^{-1}$. According to Fig.~\ref{cutsifeffbkgrej}, the optimal cuts give a slight gain for the MLP and BDT, which can exploit nonlinear correlations among $\Delta\eta_{jet}$, $\Delta\phi_{jet}$, $H_{T}$ and $\cancel{E}_{T}$. In Table~\ref{bkg_rej_sig_eff_table}, the MLP ($0.934$) and BDT ($0.933$) lead, with larger AUC once cuts are applied, while LikelihoodD ($0.870$) and Likelihood ($0.815$) trail behind. These two classifiers use the variables independently, through products of one-dimensional PDFs, which limits their ability to separate the charged Higgs signal from backgrounds with similar multi-jet and di-$\tau$ final states.

The BDT is configured with 800 trees of maximal depth 5 (Table~\ref{hyperparameters}); nodes holding less than $2.5\%$ of the events are not split further, and boosting uses AdaBoost with $\beta=0.8$. At each node the cut value is chosen by comparing the summed Gini index of the nodes produced by the split, with the range of each variable divided into 20 equal-width cells. As Figs.~\ref{cuteffbdt}, \ref{cutefflikelihood}, \ref{cutefflikelihoodd} and \ref{cuteffmlp} and Table~\ref{signalsig} show, the signal significance at $\mathcal{L}_{int}=3000~\text{fb}^{-1}$ increases once the kinematic cuts are applied, namely a higher jet $P_{T}$ threshold, a tighter $\eta_{jet}$ requirement and a higher $\cancel{E}_{T}$ threshold. These cuts reduce $\gamma\gamma\rightarrow WWZ$ and $\gamma\gamma\rightarrow t\bar{t}$, which can mimic the signal through di-$\tau$ and multi-jet final states, and improve the signal-to-background ratio in 2HDM-III charged Higgs production. We attribute the small advantage of the MLP and BDT to their ability to model nonlinear correlations among continuous variables: neural networks exploit the smooth variation of these variables to build smoother decision boundaries and thus separate events in $\gamma\gamma$ collisions more effectively.

\section{Conclusion}
We have studied $\gamma\gamma\to H^{+}H^{-}$ in the 2HDM-III for $100\leq M_{H^{\pm}}\leq190~\text{GeV}$ at a $3~\text{TeV}$ photon collider with $\mathcal{L}_{int}=3000~\text{fb}^{-1}$, using BP1 ($\tan\beta=10$, $\cos(\beta-\alpha)=-0.2$) and BP2 ($\tan\beta=20$, $\cos(\beta-\alpha)=-0.1$). The quantitative results are summarized below.

\textit{Production.} For photons radiated elastically from the lepton beams, the cross-section of both benchmark points grows with energy up to $\sqrt{s}=3~\text{TeV}$ for the $RR$, $LL$, $RL$ and unpolarized configurations; the values are displayed in the figures.

\textit{Decays.} The total width is $10^{-4}$--$10^{-1}~\text{GeV}$ over the mass range and slightly larger for BP2, so $H^{\pm}$ is a narrow resonance. In BP1 the branching ratios at $100~\text{GeV}$ ($190~\text{GeV}$) are $8.16\times10^{-1}$ ($2.81\times10^{-2}$) for $b\bar{c}$, $1.64\times10^{-1}$ ($5.65\times10^{-3}$) for $c\bar{s}$, $4.62\times10^{-3}$ ($1.79\times10^{-4}$) for $\tau\nu_{\tau}$ and $1.63\times10^{-5}$ ($6.33\times10^{-7}$) for $\mu\nu_{\mu}$. In BP2, $\mathrm{BR}(H^{+}\to\tau\nu_{\tau})$ falls from $2.91\times10^{-4}$ to $1.16\times10^{-5}$ over the same interval. The partial widths are $10^{-7}$--$10^{-5}~\text{GeV}$ for $\tau\nu_{\tau}$ and $10^{-9}$--$10^{-8}~\text{GeV}$ for $\mu\nu_{\mu}$.

\textit{Multivariate analysis.} The classifiers were trained and tested on $1.0\times10^{5}$ events per process (split $50\%/50\%$), after the preselection $P_{T}^{jet}>20~\text{GeV}$, $\eta_{jet}<2.6$, $N_{jet}\geq3$ and $\cancel{E}_{T}>50~\text{GeV}$. With cuts, the AUC values are $0.934$ (MLP), $0.933$ (BDT), $0.870$ (LikelihoodD) and $0.815$ (Likelihood), against $0.932$, $0.930$, $0.871$ and $0.814$ without cuts. For $171$ signal and $604$ background events at $M_{H^{\pm}}=190~\text{GeV}$, the maximum significances are $9.8630$, $9.8185$, $8.5369$ and $7.5528$ with cuts, against $9.78728$, $9.73028$, $8.49856$ and $7.54631$ without. The cuts therefore change the significance by $+0.77\%$ (MLP), $+0.91\%$ (BDT), $+0.45\%$ (LikelihoodD) and $+0.09\%$ (Likelihood), and the AUC by at most $0.003$. The MLP exceeds LikelihoodD by $1.33$ ($15.5\%$) and Likelihood by $2.31$ ($30.6\%$) in significance.

\textit{Reach and limitations.} The MLP significance of $9.86$ corresponds to $1.97$ times the conventional threshold of $5$, counting statistical fluctuations only, and the significance grows with both $M_{H^{\pm}}$ and $\mathcal{L}_{int}$ (Fig.~\ref{significance}). The classifier results above are quoted for $M_{H^{\pm}}=190~\text{GeV}$. Systematic uncertainties and a quantitative comparison with LHC limits are not included, so the reach at other masses and relative to the LHC remains to be quantified.




\end{document}